# Determination of the Total Solar Modulation Factors in the Heliosphere For Cosmic Ray Protons and Electrons by Comparing Interstellar Spectra Deduced from Voyager Measurements and PAMELA Spectra of These Particles at the Earth

W.R. Webber

1. New Mexico State University, Department of Astronomy, Las Cruces, NM 88003, USA

## ABSTRACT

We have determined the interstellar spectra of cosmic ray protons and electrons from a few MeV to ~10 GeV.  These interstellar spectra are based on Voyager data and a normalization of specific galactic propagation model calculations of both protons and electrons to PAMELA data at the Earth at 10 GeV, where the solar modulation is small.  These resulting interstellar spectra are then compared with spectra of protons and electrons measured at lower energies at the Earth by PAMELA in 2009.  The total amount of modulation at lower rigidities (energies) is found to be nearly the same at the same rigidity for both protons and electrons and ranges in magnitude from a factor ~400 at 0.1 GV for electrons, to a factor ~15 at 0.44 GV (100 MeV for protons), to a factor ~3.3 at 1 GV for both components.  The magnitude of this total modulation of both components are the same to within $\pm$ 10% from ~0.3 to ~3 GV in rigidity.  The observed total modulation for protons can be matched quite closely using a simple spherically symmetric modulation picture involving a force field model for the modulation and a constant energy loss at all energies.  The electrons require a set of parameters to calculate more detailed features of the modulation using the diffusion coefficient and its variation with rigidity and radius at rigidities less than few GV.

**Introduction**

The overall effects of solar modulation in the heliosphere beyond the Earth have hidden the true interstellar spectrum of galactic cosmic rays below a few GeV since their discovery as extra solar particles over 60 years ago. This vail has finally been lifted by the measurements of Voyager 1 (V1) galactic cosmic rays beyond the heliopause starting in 2012 (Stone, et al., 2013; Webber and McDonald, 2013). Electrons from 5-60 MeV, protons and helium nuclei from ~3-600 MeV and the spectra of heavier nuclei are Voyagers on-going legacy. It is fortunate that, at roughly the same time, extremely accurate measurements were being made of these cosmic ray protons and electrons between ~100 MeV and 100 GeV, by the PAMELA spacecraft (Adriani, et al., 2013, 2015), at a time in 2009 when the solar modulation affects were at a historical minimum and the intensities at the Earth were at a maximum.

In this paper we will compare the local interstellar spectra (LIS) measurements and calculations of protons and electrons beyond the heliopause and the PAMELA measurements of these same particles at the Earth, thus obtaining an overview of the magnitude of this modulation between 2009-2013 from a few MeV to 10 GeV for both protons and electrons.

For electrons the total modulation factor between PAMELA and Voyager at ~100 MeV is found to be a factor ~400, whereas at 1 GeV this modulation is only a factor of between 3 and 4. For protons this total modulation factor is ~15 at 100 MeV, decreasing to a factor ~1.8 at 1 GeV. These ratios are in 2009 when the cosmic ray intensity levels were at a historical high at the Earth.

These total modulation factors are of great interest in a wide variety of studies involving the production of radioactive nuclei at the Earth and throughout the solar system, for example, Cosmogenic Radio Nuclei (Beer, McCracken and von Steiger, 2012).

**The Data**

The intensities for protons at V1 (Cummings, et al., 2015) and at PAMELA in 2009, (Adriani, et al., 2013) along with the derived interstellar galactic proton spectrum is shown in Figure 1.

The intensities for electrons at V1 (Cummings, et al., 2015, 2016) and at PAMELA in 2009, (Adriani, et al., 2015) along with the derived interstellar electron spectrum is shown in Figure 2. Figure 2 is j x $E^2$.

The calculation of the interstellar proton and electron spectra that match the V1 intensities at lower energies and the PAMELA intensities at 10 GeV is described in the following section.

## The Calculation of the Interstellar Proton and Electron Spectra

As can be seen in Figures 1 and 2 the spectra measured at V1 and at PAMELA do not overlap well in energy. However in each case we can calculate an interstellar spectrum that connects the V1 data to the PAMELA data at high energies at 10 GeV/nuc, where the modulation is small. To obtain the proton spectra in Figure 1 we have used a Leaky Box model for galactic propagation similar to that used by Webber and Higbie, 2008, 2009, 2013; Webber, 2015. The Webber and Higbie, 2009, version of this model has already provided an accurate prediction of the subsequent Voyager measurements of the interstellar spectra of H and He nuclei above 80 MeV/nuc (see Cummings, et al., 2015). The important parameters used in the model are 1) The mean free path, defined by the B/C ratio and taken to be $\lambda$=22.3 $\beta$ $P^{-0.47}$ (g/cm$^2$) above a rigidity $P_0$. Below $P_0$, which is taken to be 1.0 GV, the interstellar diffusion coefficient changes by one power in the exponent from being ~$P^{0.47}$ at high rigidities to $P^{-0.53}$ at low rigidities (see discussion by Ptuskin, et al., 2006); 2) The source spectra of protons is taken to have an exponent of -2.26 which is constant as a function of rigidity for the calculations.

Specifically for this calculation we normalize the propagated proton intensity to that measured by PAMELA at 10 GeV + 10% (to allow for solar modulation) = 230 p/m$^2$ sr·s·GeV. The diffusion coefficient is taken to be proportional to $P^{0.47}$ above a rigidity $P_0$. Below that rigidity, K~$P^{-0.53}$, a change of 1.0 power in the exponent occurs. The changing exponent of the resulting propagated proton spectrum, which is easily seen in Figure 1, is due purely to propagation effects with no need to postulate different source spectral indices at high and low energies.

Consider now the electron spectra shown in Figure 2. The propagated interstellar spectrum is calculated using a Monte Carlo Diffusion model that has been used several times in the past for calculating the interstellar electron spectrum (e.g., Webber and Rockstroh, 1997; Webber and Higbie, 2008, 2013; Webber, 2015b). The important parameters are: 1) the lifetime, $\tau = L_z^2/2K$, where $L_z = 1.5$ kpc which is the height of the diffusing region taken to be a disk above galactic plane, K equals the diffusion coefficient $= 2 \times 10^{28}$ cm$^2$/s at 1 GeV. The diffusion lifetime at 1 GeV therefore becomes $1.5 \times 10^7$ yr. which is approximately that measured by Yanasuk, et al., 2001, and; 2) The source spectrum for electrons, here also taken to be a constant $= -2.25$ at all energies (rigidities).

This electron calculation includes the energy loss from synchrotron radiation where B=5μG and also inverse Compton loss which becomes important above a few GeV. These terms exceed the diffusion loss term in magnitude at energies greater than a few GeV and are $\sim E^2$ and are thus responsible for the changing spectral index above 1 GeV that is seen in Figure 2. Ionization loss and bremsstrahlung loss at lower energies in a matter disk of variable thickness and density are also included (see earlier references for details).

Specifically, what is new for this calculation beyond the earlier calculations is that we normalize the propagated electron intensity to be equal to the PAMELA 2009 electron intensity measured at 10 GeV+10%=2.4 e/m$^2$·s·sr·GeV. This extra 10% above the measurements is due to the estimated residual solar modulation in 2009. We also consider the diffusion coefficient to be $\sim P^{0.47}$ above a rigidity $P_0$. Below that rigidity, which is taken to be 1 GV, K $\sim P^{-0.53}$, which is a change of 1.0 power in the rigidity dependence of the diffusion coefficient. These dependences are consistent with new measurements of the B/C ratio by AMS-2 (Oliva, CERN, 2015) which define this ratio to within a few % between 1 and 800 GeV/nuc and with new fits to the B/C data measured by Voyager (Cummings, et al., 2015; Webber and Villa, 2016).

The new electron calculation is extended from a lower limit of 13 MeV in the original models down to 0.5 MeV. This is in response to new Voyager measurements of the LIS spectrum extending to below 3 MeV (Cummings, et al., 2016). This propagated spectrum is for a source of electrons with a constant radial intensity at Z=0 in the galaxy. The new V1 spectra are shown as a red line below 10 MeV in Figure 2.

**Comparison of the Proton/Electron Spectra at V1 and PAMELA - The Overall Heliospheric Solar Modulation in 2009**

In Figure 3 we show the ratios of the calculated proton and electron interstellar intensities, including also those measured at Voyager, to the PAMELA intensities measured at the Earth in 2009 shown in Figures 1 and 2. This ratio is shown on a vertical scale as $\ell$n (V1/PAM) vs. particle rigidity. The overall magnitude of the modulation of protons and electrons is indistinguishable to a level ±10% for the rigidities considered here and for the time period in 2009 when the comparison is made.

**A Description of the Solar Modulation for Protons**

The results shown in Figure 3 may be used to describe the overall effects of the solar modulation of protons in the heliosphere in a very simple way. In fact, the original models of the modulation for protons and other nuclei in the heliosphere were framed as early as 1957 in terms of Louvilles Theorem and how the input spectrum of protons and heavier nuclei (the interstellar spectra) would change if all of the particles lost an equal amount of energy in the heliospheric modulation process (Ehmert, 1957). Later Gleeson and Axford, 1968, by making various approximations to a spherically symmetric transport diffusion equation for particles in the heliosphere, derived a force-field solution in which an energy loss appears in the solution. Both approaches give

$$j(E) = j(E+\Phi) \cdot [E(E+2E) / (E + \Phi)(E + \Phi + 2E_0)]$$

where E is the energy at the Earth, $\Phi$ is the amount of energy loss (potential difference in MV), $E_0$ is the rest mass energy/nuc = 931 MeV and j(E) is the intensity at the Earth.

We have used this simple formula to calculate the ratio of the V1 (≡LIS) proton spectrum to that observed at PAMELA for two examples, for energy losses of 250 and 300 MV. These calculations are shown in Figure 4, superimposed on the observed proton modulation. The calculations bracket the observed ratios for protons quite closely over a rigidity range from ~100 MV to several GV. At 1 GV the E loss would be ~285 MV. On the whole this simple force field formulation does a remarkable job of describing the overall heliospheric proton modulation thus justifying its widespread use in the literature (McCracken, Chapter 5.7, pp 44-63 in Cosmogenic Radionuclides).

Mewaldt, et al., 2011, have noted, based on measurements on the ACE spacecraft and other measurements, that the highest intensities in the modern era were observed in 2009 and corresponded to a solar modulation level ~250-270 MV. Lave, et al., 2013, used the intensities of C and heavier nuclei measured on ACE to trace the solar modulation level throughout a full solar cycle and found a minimum modulation in 2009 corresponding to ~250 MV. Both of these modulation estimates were based on more sophisticated, but still spherically symmetric force field equivalent modulation models.

Usoskin, et al., 2012, have used neutron monitor data to trace the solar modulation level over several solar cycles, obtaining a minimum modulation ~250 MV during 2009.

The relative solar modulation of positive and negative particles has also been an issue, as discussed in Potgieter, 2014. The results of this paper are compatible with an indistinguishable difference in the total overall solar modulation of the two components at a level ~ $\pm$ 10%. This new measurement needs to be reconciled with "local" measurements near the Earth and inside the termination shock which clearly show a different modulation for protons and electrons, in some cases with a general 11-22 year solar periodicity, (e.g., Webber, Heber and Lockwood, 2005; Potgieter, et al., 2015).

## <u>Summary and Conclusions</u>

This paper presents results on the total solar modulation of protons and electrons in the heliosphere as determined from the spectra in local interstellar space measured by Voyager or deduced by galactic propagation models and the spectra at Earth as measured by the PAMELA spacecraft. These interstellar spectra are normalized to the PAMELA intensities measured at 10 GeV where the solar modulation is small, so that the relative LIS spectra and Earth-based spectra can be obtained between ~100 MeV and 10 GeV.

The total amount of modulation between interstellar space and the Earth in 2009 is found to be the nearly same at the same rigidity for both protons and electrons in overlapping rigidity ranges. This modulation ranges from a factor ~350 at 0.1 GV (for electrons) to a factor ~15 at 0.44 GV (for 100 MeV protons) to a factor of ~3.3 at 1.0 GV (for both components) decreasing to ~10% at 10 GV.

This measured rigidity dependence and the magnitude of the total modulation are the same within $\pm$ 10% for electrons and protons between ~0.3 – 3.0 GV, implying that there is no significant charge dependence of the overall solar modulation observed a the location of the Sun at this time.

The total modulation for protons can be matched quite closely by a simple force field model which considers a constant energy loss at all energies. This modulation can be described in terms of potential energy loss ~285 MV at 1 GV at the time of the PAMELA measurement in 2009. For A/Z nuclei = 2.0, this loss is 285/2 = 142.5 MV/nuc. For the electrons, the calculation is more complicated and below a few GV it depends on the value of effective diffusion coefficient and its rigidity dependence inside and outside of the heliospheric termination shock. This is a subject for a future paper using the radial dependence of Voyager electron spectra that are observed beyond the heliospheric termination shock where most of the electron modulation may occur.

**Acknowledgements:** The author appreciates the efforts of all of the Voyager CRS team, Ed Stone, Alan Cummings, Nand Lal and Bryant Heikkila. It is a pleasure to work with them as we try to understand and interpret the galactic cosmic ray spectra that had previously been hidden from us by the solar modulation. We also thank JPL for their continuing support of this program for over 40 years. This work would not have been possible without the assistance of Tina Villa.

# Figure Captions

**Figure 1:**  Proton intensity measurements from ~2 MeV to 10 GeV.  Red points are Voyager data, blue points are PAMELA data in 2009.  Black curve is the propagated galactic proton spectrum (see text) with a source spectrum with index S = -2.26 and with this index independent of rigidity and normalized to the PAMELA intensity measured at 10 GeV plus 10% for solar modulation.

**Figure 2:**  Electron intensity measurements from ~3 MeV to 40 GeV.  Intensities are x $E^2$ (in GeV).  Red points and circles are Voyager data, blue points are PAMELA data in 2009. Black curve is propagated galactic electron spectrum with a spectral index S = -2.25, independent of energy and normalized to the PAMELA intensity measured at 10 GeV plus 10% for solar modulation.  The red solid curve is an extension of the Webber and Higbie, 2009, calculations to energies below 13 MeV.

**Figure 3:**  Ratio of local interstellar spectra to those measured in 2009 at PAMELA, $\ell$n ($j_V/j_P$). Black points are for electrons from 80 MeV to 4 GeV as obtained from Figure 2.  Red circles are for protons from 60 MeV to 4 GeV as obtained from Figure 1. These points above are all based on PAMELA energies.  Solid red points are a direct comparison between Voyager and PAMELA proton measurements in the energy range 80-350 MeV.  Open circles are a comparison of PAMELA data with propagated galactic spectrum.

**Figure 4:**  A comparison of the proton total modulation $\ell$n ($j_V/j_P$) shown in Figure 3 as red solid or open circles with calculations using a simple spherically symmetric force field modulation model for the heliosphere with an energy loss between 250 or 300 MV, independent of energy in the heliosphere.

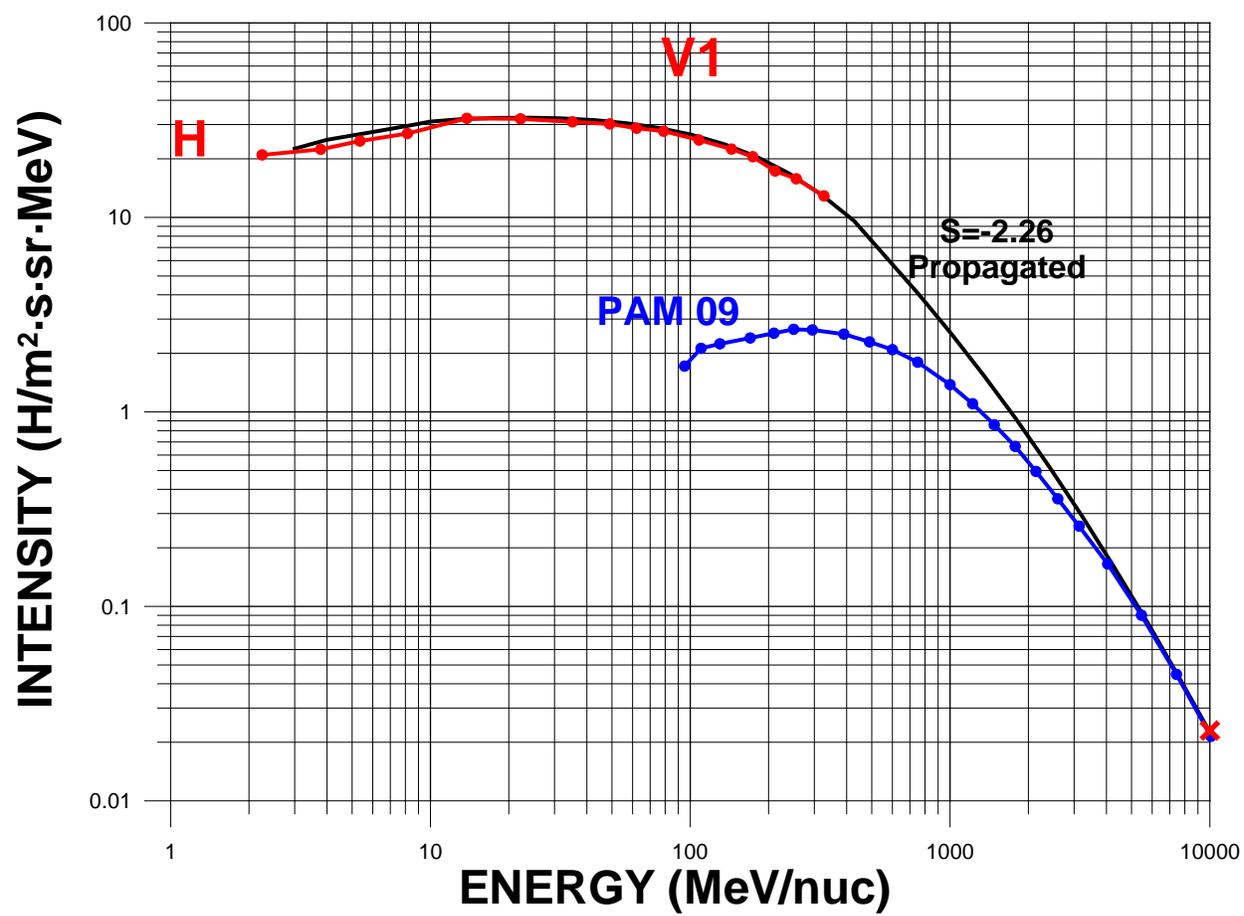

**FIGURE 1**

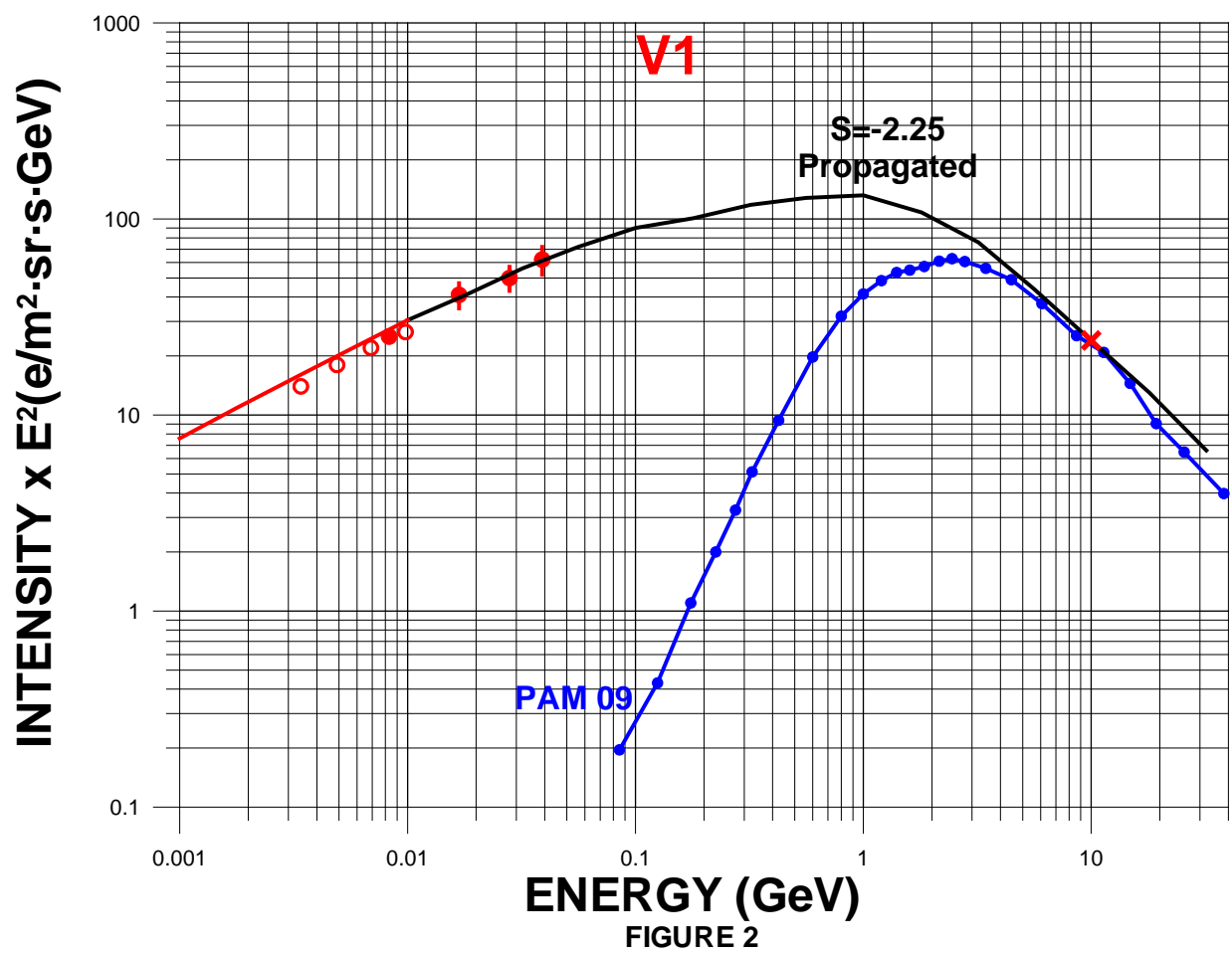

FIGURE 2

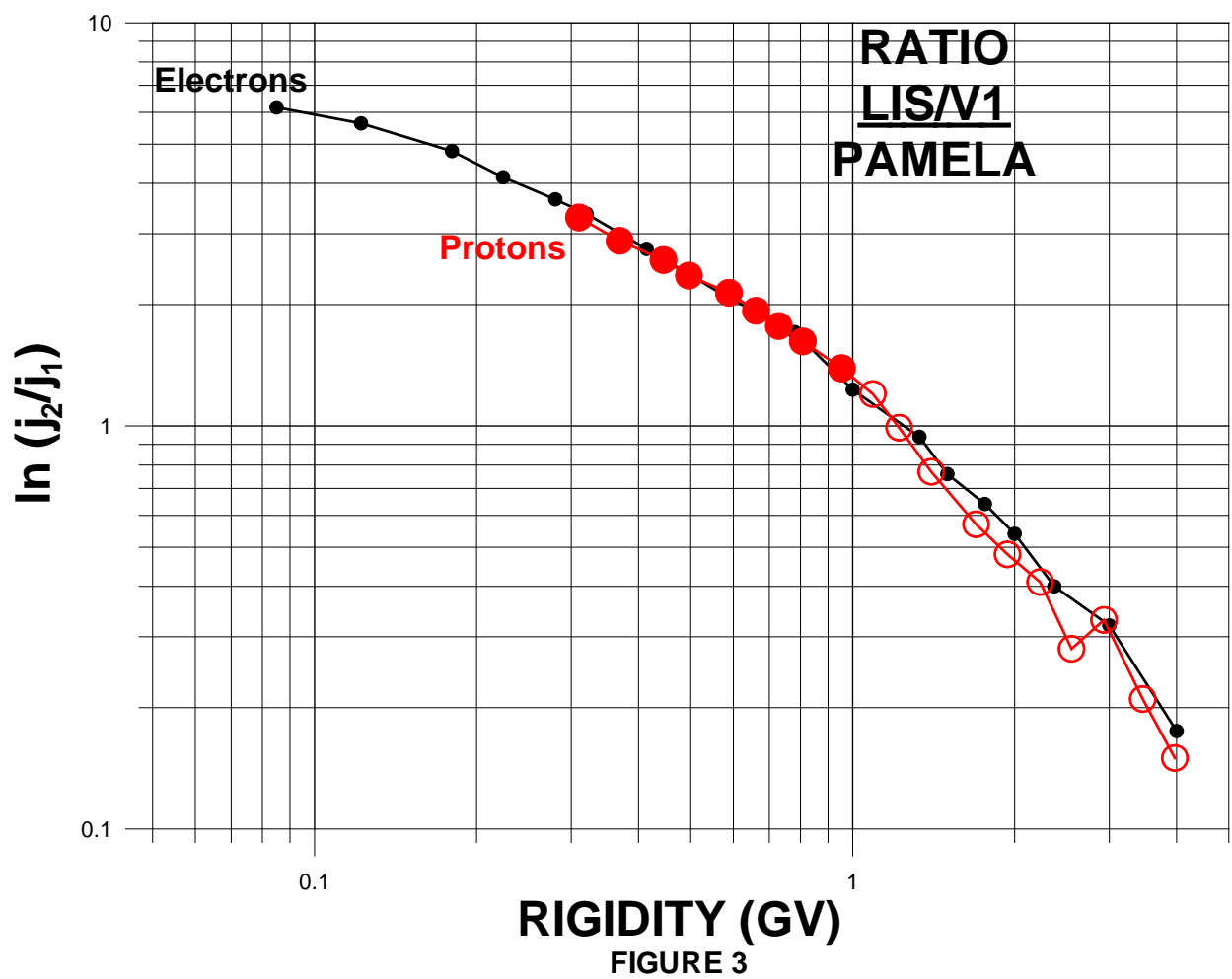

FIGURE 3

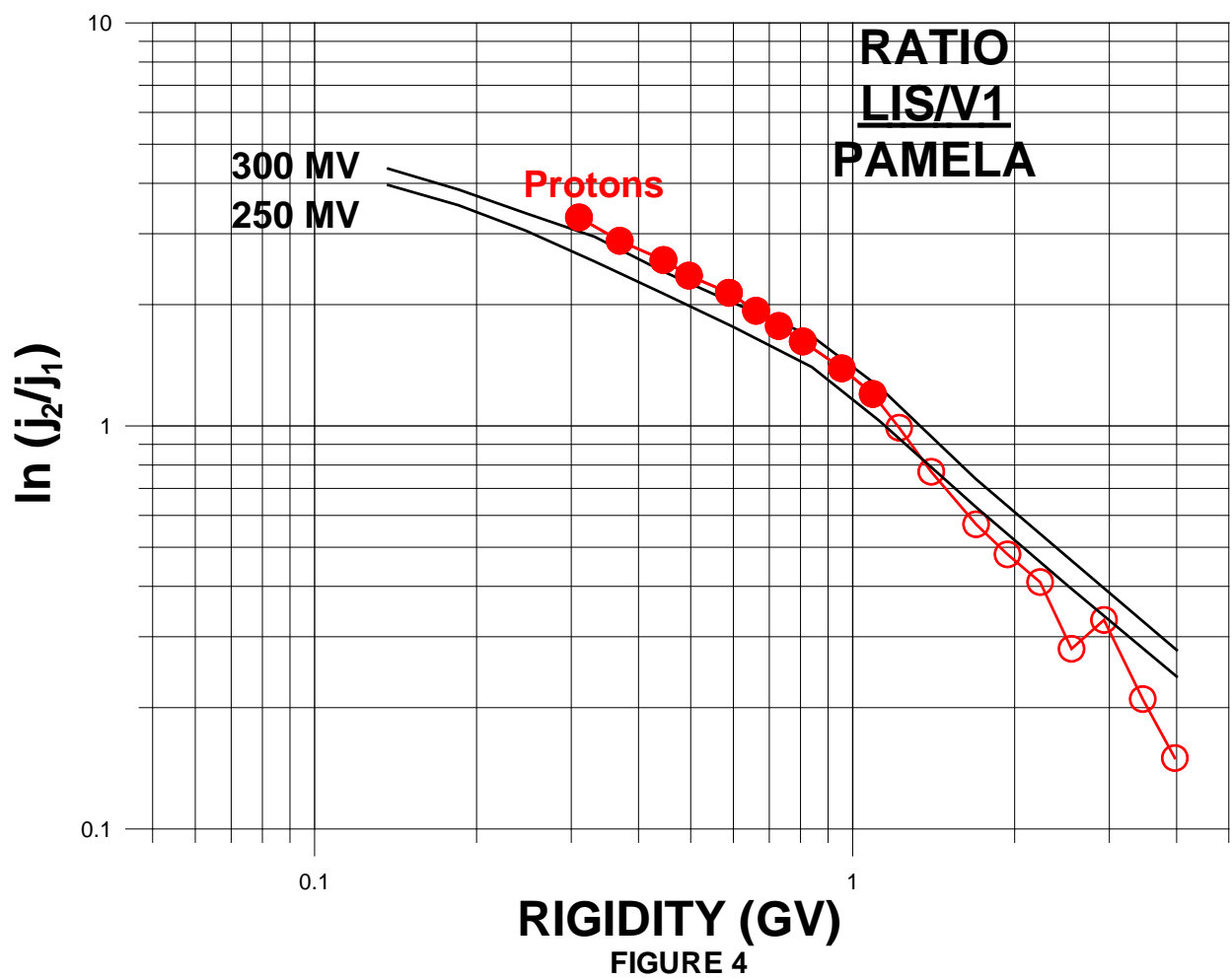

FIGURE 4